\documentclass[cits]{PoS}
\pdfoutput=1
\usepackage{rotating,booktabs,amsmath,amssymb, multicol,url}
\usepackage[utf8]{inputenc}
\makeatletter
     {\bgroup\raggedright\small\section*{\refname
        \@mkboth{\MakeUppercase\refname}{\MakeUppercase\refname}}%
      \list{\name{bib\@arabic\c@enumiv}% HOPE!
	    \@biblabel{\@arabic\c@enumiv}}%
           {\settowidth\labelwidth{\@biblabel{#1}}%
            \setlength\itemsep{0pt}
            \setlength\parskip{0pt}
            \setlength\parsep{0pt}
            \setlength\partopsep{0pt}
            \leftmargin\labelwidth
            \advance\leftmargin\labelsep
            \@openbib@code
            \usecounter{enumiv}%
            \let\p@enumiv\@empty
            \renewcommand\theenumiv{\@arabic\c@enumiv}}%
      \sloppy\clubpenalty4000\widowpenalty4000%
      \sfcode`\.\@m}
     {\def\@noitemerr
       {\@latex@warning{Empty `thebibliography' environment}}%
      \endlist\egroup}
\makeatother

\newcommand{\gGF}{\ensuremath{g_{\rm GF}^2} }

\newcommand{\MSbar}{\ensuremath{\overline{\textrm{MS}} } }

\newcommand{\vev}[1]{\ensuremath{\left\langle #1 \right\rangle} }

\newcommand{\fig}[1]{Fig.~\ref{#1}}

\title{Continuous $\beta$ function for the SU(3) gauge systems with two and twelve fundamental flavors}

\ShortTitle{Continuous $\beta$ function}

\author{\speaker{Anna Hasenfratz}\\
        Department of Physics, University of Colorado, Boulder, CO 80309, USA\\
        E-mail: \email{anna.hasenfratz@colorado.edu}}
\author{Oliver Witzel\\
        Department of Physics, University of Colorado, Boulder, CO 80309, USA\\
        E-mail: \email{oliver.witzel@colorado.edu}}

\abstract{
The gradient flow transformation can be interpreted as continuous real-space renormalization group transformation if a coarse-graining step is incorporated  as part of calculating expectation values.  The method allows to predict critical properties of strongly coupled systems including the renormalization group $\beta$ function and anomalous dimensions at nonperturbative fixed points. In this contribution we discuss a new analysis of the continuous renormalization group  $\beta$ function  for $N_f=2$ and $N_f=12$ fundamental flavors in SU(3) gauge theories based on this method. We follow the approach developed and tested  for the $N_f=2$ system in   arXiv:1910.06408.   Here we  present further information on the analysis,  emphasizing the robustness and intuitive features of the continuous $\beta$ function calculation.    
We also discuss the applicability of the continuous $\beta$ function calculation in conformal systems, extending the possible phase diagram to include a 4-fermion interaction. The numerical analysis for $N_f=12$ uses the same set of ensembles that was generated and analyzed for the step scaling function in arXiv:1909.05842. The new analysis uses volumes with $L \ge 20$ and determines the $\beta$ function in the $c=0$ gradient flow renormalization scheme. The continuous $\beta$ function predicts the existence of a conformal fixed point and is consistent between different operators.  Although determinations of the step scaling and continuous $\beta$ function  use different renormalization schemes, they both predict the existence of a conformal fixed point around $g^2\sim 6$.}
  
\FullConference{37th International Symposium on Lattice Field Theory - Lattice2019\\
		16-22 June 2019\\
		Wuhan, China}

\begin{document}

\section{Introduction}

Real-space renormalization group (RG) flows map out the phase structure of lattice models  including properties of  infrared  and ultraviolet  fixed points (IRFP, UVFP). 
Recently an interpretation of gradient flow (GF) transformations as a continuous real-space RG transformation has been proposed \cite{Carosso:2018bmz}. This opens a new way to determine the RG $\beta$ function and anomalous dimensions of strongly coupled gauge-fermion systems both within and outside the conformal window. In Reference \cite{Hasenfratz:2019hpg} we developed the steps to determine the continuous RG $\beta$ function of the GF coupling   illustrating the method for the QCD-like SU(3)  model with 2 fundamental flavors. 
The new approach has several advantages compared to commonly performed step-scaling calculations. It is equally applicable for confining, conformal, or infrared free systems. 
In this work we extend the calculation of Ref.~\cite{Hasenfratz:2019hpg} to 12 fundamental flavors.

\begin{figure}[tb]
\centering
\includegraphics[width=0.475\columnwidth]{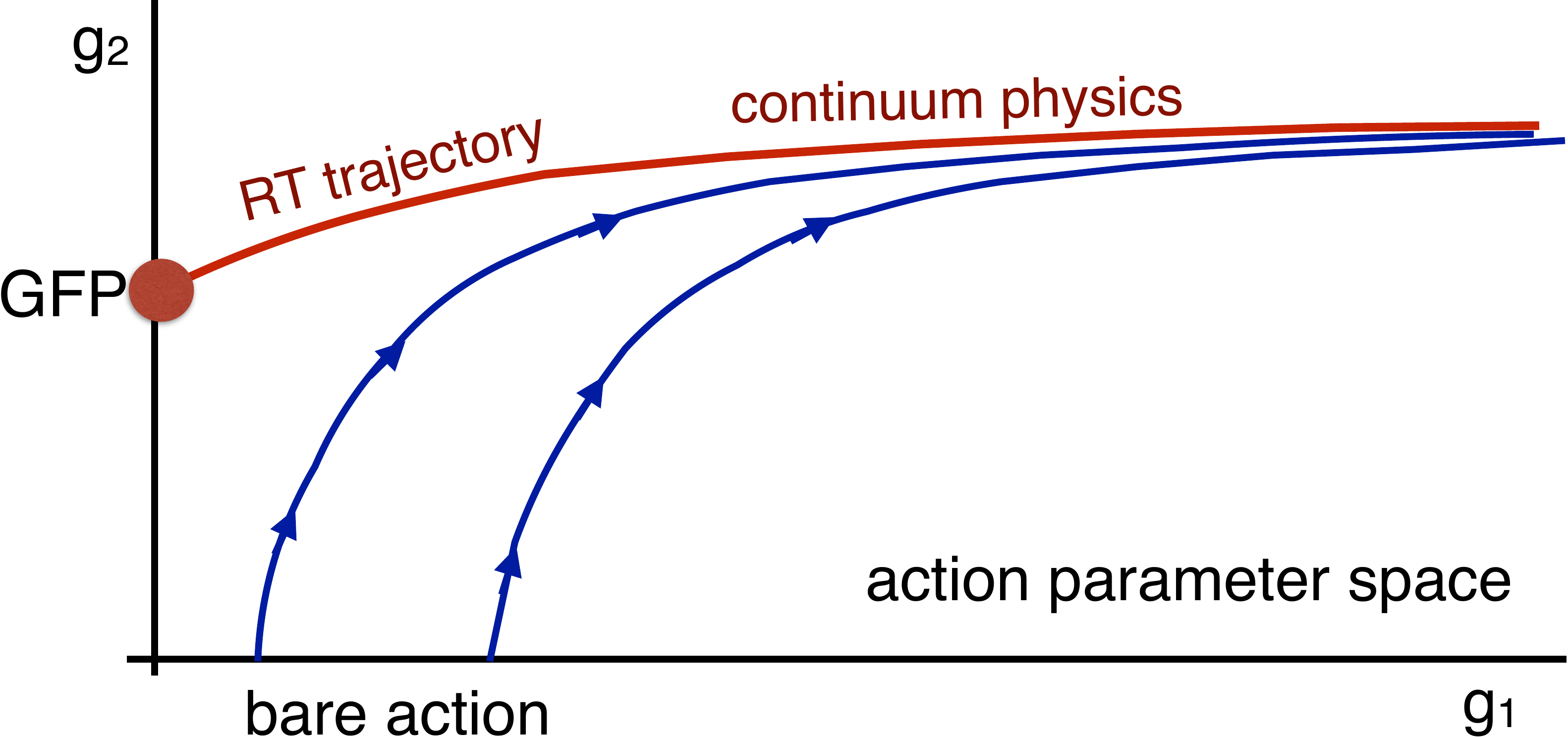} \hfill
\includegraphics[width=0.495\columnwidth]{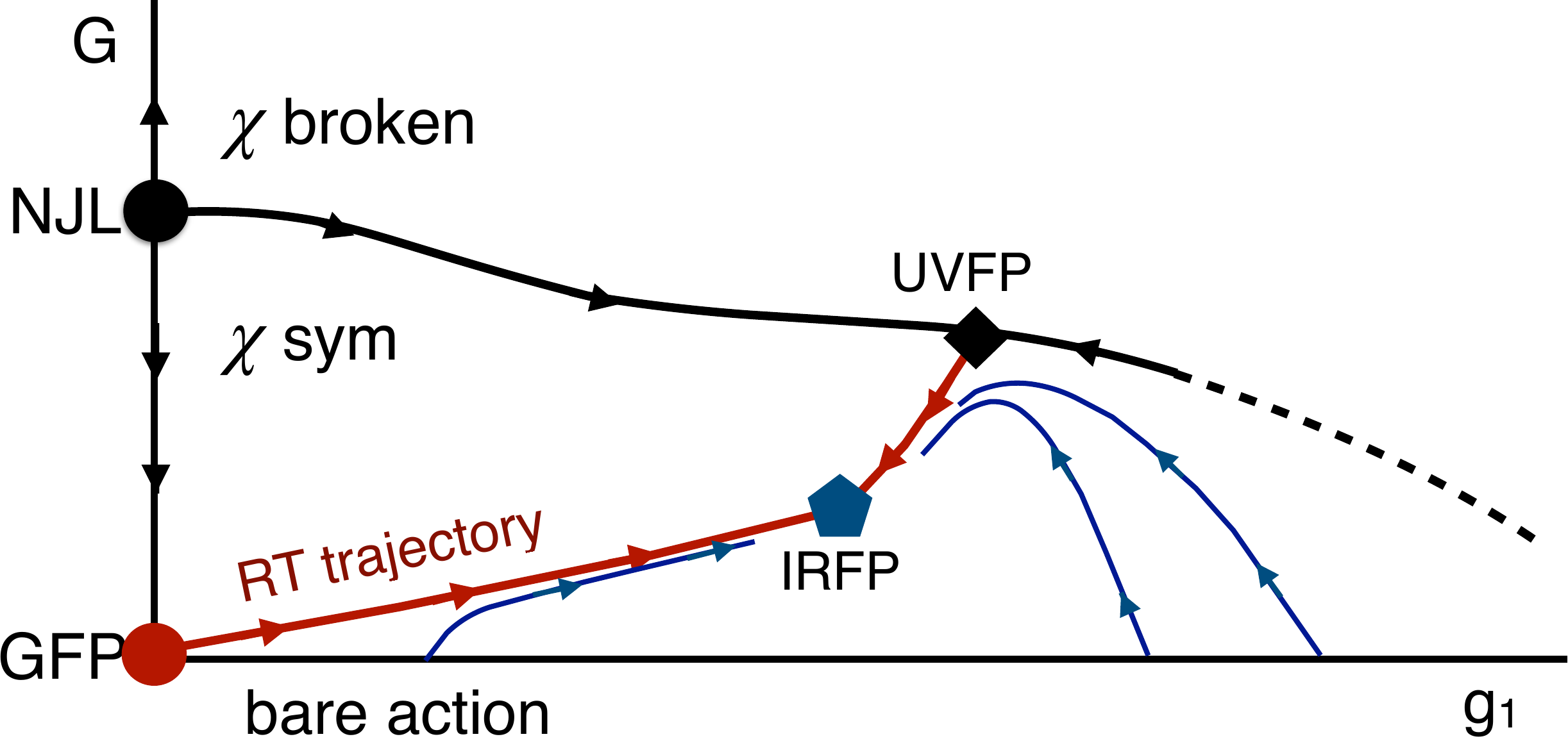}
\caption{Sketch of possible phase diagrams and RG flows  in the multi-dimensional action parameter space. Left panel: QCD-like gauge-fermion system. $g_1$ refers to the relevant gauge coupling, while  $g_2$ indicates all other irrelevant couplings. The blue lines represent RG flow trajectories. Right panel: Infrared conformal system with 4-fermion interaction. $G$ denotes the 4-fermion coupling and only the relevant $g_1$ coupling of the gauge-fermion system is shown. The black  solid line denotes a 2nd order phase transition separating chirally symmetric and broken phases. The black dashed line indicates that the phase transition might turn 1st order but it remains phase separating. The RT emerging from the GFP  and from a UVFP on the phase separating surface end at an IRFP. RG flows in the weak coupling side are similar to those on the left panel but can become complicated on the strong coupling side of the IRFP. }
\label{fig:RG-flow}
\end{figure}

 GF is a continuous  transformation that can be used to define real-space RG blocked quantities. The coarse graining of the RG transformation can be incorporated when calculating expectation values \cite{Carosso:2018bmz}.   Relating the dimensionless GF time $t/a^2$  to the RG scale change  as $b \propto \sqrt{t/a^2}$, the GF transformation describes  a continuous real-space RG transformation.  In particular, expectation values of local operators, like the energy density, are identical with or without coarse graining and describe at flow time $t/a^2$  physical quantities at energy scale $\mu \propto 1/\sqrt{t}$.
We sketch typical RG flows on the chiral $m=0$ critical surface in an asymptotically free gauge-fermion system   in the left panel of  \fig{fig:RG-flow}. $g_1$ refers to the relevant gauge coupling, while  $g_2$ indicates all other irrelevant couplings. The Gaussian FP (GFP) and the renormalized trajectory (RT) emerging from it describe the cut-off independent continuum limit. 
The RT is a 1-dimensional line.  A dimensionless  local operator with non-vanishing expectation value can therefore be used to define a running coupling along the RT. The simplest such quantity in gauge-fermion systems is $\vev{t^2 E(t)}$, the energy density multiplied by $t^2\propto b^4$. This is   the quantity proposed in Ref.~\cite{Luscher:2010iy} to define the gradient flow renormalized coupling $\gGF$.
Numerical simulations are performed with an action characterized by a set of bare couplings. If this action is in the vicinity of the GFP or its RT, the typical RG flow approaches the RT and follows it as the energy scale is decreased from the cut-off towards the infrared as indicated by the blue lines. RG flows starting at  different   bare couplings   approach the RT differently.  However, once the irrelevant couplings have died out, they all follow the same 1-dimensional renormalized trajectory and describe the same continuum physics.
 At large flow time, irrelevant terms in the lattice definition of $E(t)$ die out as well and $\gGF$ approaches a continuum renormalized running coupling. Its derivative is the RG $\beta$ function
 \begin{equation}
\beta(g_{GF}^2) = \mu^2 \frac{d g_{GF}^2}{d \mu^2} = -  t \frac{d g_{GF}^2}{d t}.
\label{eq:contbfn}
\end{equation}

The RT of chirally broken systems continues to $g_1 \to \infty$,  whereas conformal systems have an IRFP on the RT that stops the flows from either direction. 
It was suggested that the IRFP of conformal systems is accompanied by a UVFP~\cite{Miransky:1998dh,Kaplan:2009kr,Gorbenko:2018ncu}. The latter requires the emergence of a new relevant operator, possibly a 4-fermion interaction. In the right panel of \fig{fig:RG-flow}, we sketch a possible phase diagram for this case\footnote{The first version of this phase diagram emerged in a discussion between Slava Rychkov and A.~H. during the TASI Summer School in June 2019.}. We show only the relevant $g_1$ coupling of the gauge-fermion system  and include $G$, the coupling of the 4-fermion Nambu-Jona-Lasinio (NJL) interaction~\cite{NJL1,NJL2}.  The NJL model has a second order phase transition separating chirally symmetric and broken phases at $g_1$=0. Lattice studies indicate that  for $g_1>0$ at least initially the phase transition remains continuous  \cite{Catterall:2013koa,Rantaharju:2016jxy,Rantaharju:2017eej}.  At some point it might turn first order but the phase transition must persist as it separates distinct phases.  A UVFP could sit on the continuous region of this phase separating surface and be connected to the IRFP as indicated. Since the phase transition of the un-gauged NJL model does not depend on the number of fermions, it is conceivable that the phase separating surface remains far from the $G=0$ gauge fermion system. This might imply a sharp turn of the RT making the numerical study of the ``backward flow'' particularly challenging.\\

 The above discussion and the definition of the $\beta$ function in Eq.~(\ref{eq:contbfn}) is valid in infinite volume.  In our approach we extrapolate  $L/a \to \infty$  at fixed $t/a^2$ which  also sets the renormalization scheme  $c=\sqrt{8t}/L = 0$. 
The continuum limit of the $\beta$ function is obtained at fixed $\gGF$ while taking $t/a^2\to\infty$. In QCD-like systems this automatically forces the bare gauge coupling towards zero, the  critical surface of the GFP. In a conformal system the bare gauge coupling is tuned to zero in the weak coupling regime, whereas in the strong coupling ``backward flow'' regime it should be tuned to the phase separating critical surface. 
Specifically, once the GF coupling and its derivative are determined, the continuous $\beta$ function calculation proceeds in two steps:
\begin{itemize}
\item[A)]  Infinite volume extrapolation at every GF time: the leading order  corrections at small GF values are $(a/L)^4$. We restrict the flow time such that the finite volume corrections are well described by the leading behavior.
\item[B)] Infinite flow time extrapolation at every $g^2_{GF}$: this step removes irrelevant operator contributions  (cut-off effects) and plays the role of the $a/L \to 0$ continuum limit extrapolation of the step-scaling function approach. 
\end{itemize}
Step A)  is  new in the continuous $\beta$ function approach but is compensated  by other advantages. Most importantly the flow time in the continuous $\beta$ function calculation is independent of the volume and can be kept small. This significantly reduces  statistical errors.

\section{Numerical details}
Our lattice  studies of both 2 and 12-flavor SU(3) systems  are based on gauge field configurations generated with tree-level improved Symanzik gauge action and chirally symmetric M\"obius domain wall (DW) fermions with stout smeared gauge links. We generate configurations using \texttt{Grid} \cite{Boyle:2015tjk,GRID} with $a m=0$ bare mass and chose the DW 5th dimension large enough to ensure that the residual mass is $a m_\text{res}<10^{-5}$.  Details of the $N_f=2$ configurations are discussed in Ref.~\cite{Hasenfratz:2019hpg}. The $N_f=12$ analysis uses configurations generated for the step-scaling study published in Refs.~\cite{Hasenfratz:2017qyr,Hasenfratz:2019dpr}. We have implemented three different flows, Wilson (W), Symanzik (S), and Zeuthen (Z), in \texttt{Qlua} \cite{Pochinsky:2008zz,qlua} and three different operators Wilson plaquette (W), clover (C),  and Symanzik (S) to estimate the energy density \cite{Ramos:2014kka,Ramos:2015baa}.
Our data analysis is performed using the $\Gamma$-method \cite{Wolff:2003sm} which is designed to estimate and account for autocorrelations.

The GF coupling is defined as
\begin{align}
    \label{eq:pert_g2}
    \gGF(t; L,g^2_0) = \frac{128\pi^2}{3(N^2 - 1)} \frac{1}{1+\delta} \vev{t^2 E(t)}\,.
\end{align}
The normalization  ensures that $\gGF$ matches the $\MSbar$ coupling at tree level, and the term $1/(1+\delta)$  corrects for the gauge zero modes due to periodic gauge boundary conditions~\cite{Fodor:2012td}.\footnote{$\delta$ depends on the flow time and the aspect ratio of the lattice volume. We thank D. Nogradi for sharing with us his results on the latter prior publication.}

 \begin{figure}[tb]
\centering
\includegraphics[width=0.494\columnwidth]{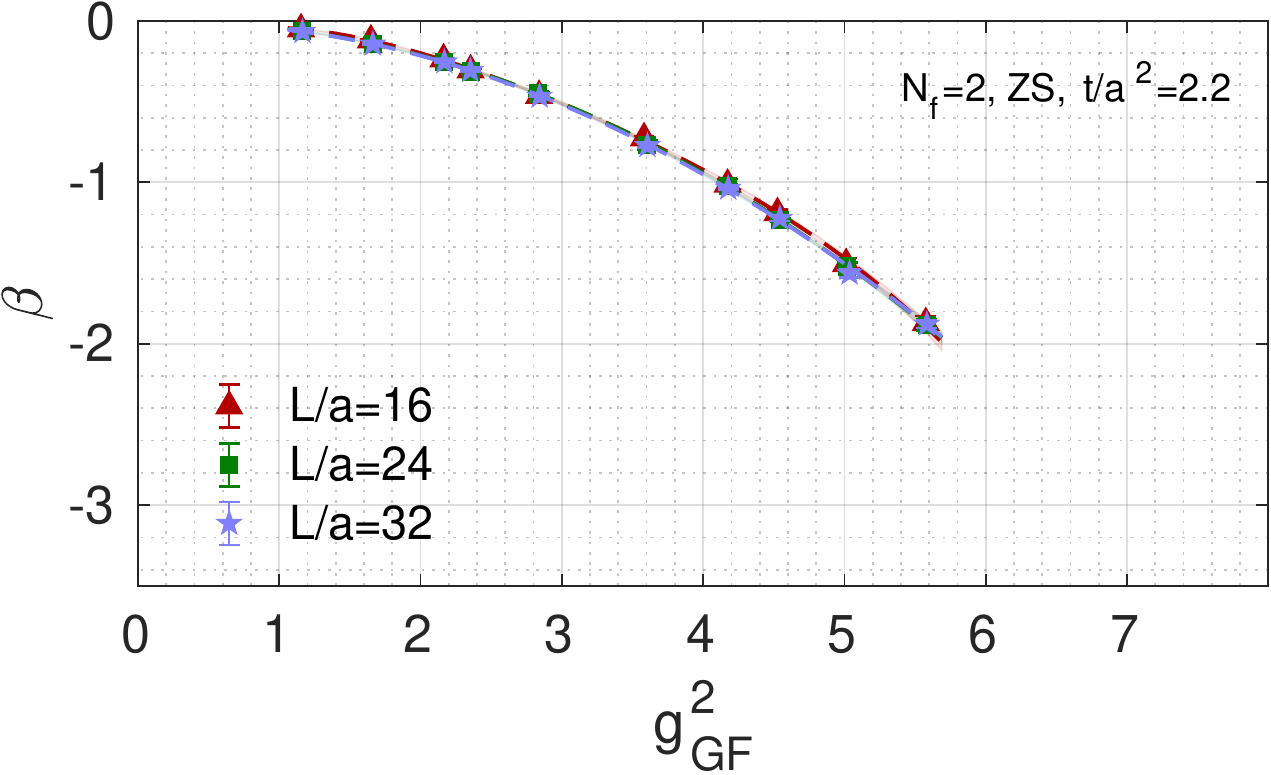}
\includegraphics[width=0.494\columnwidth]{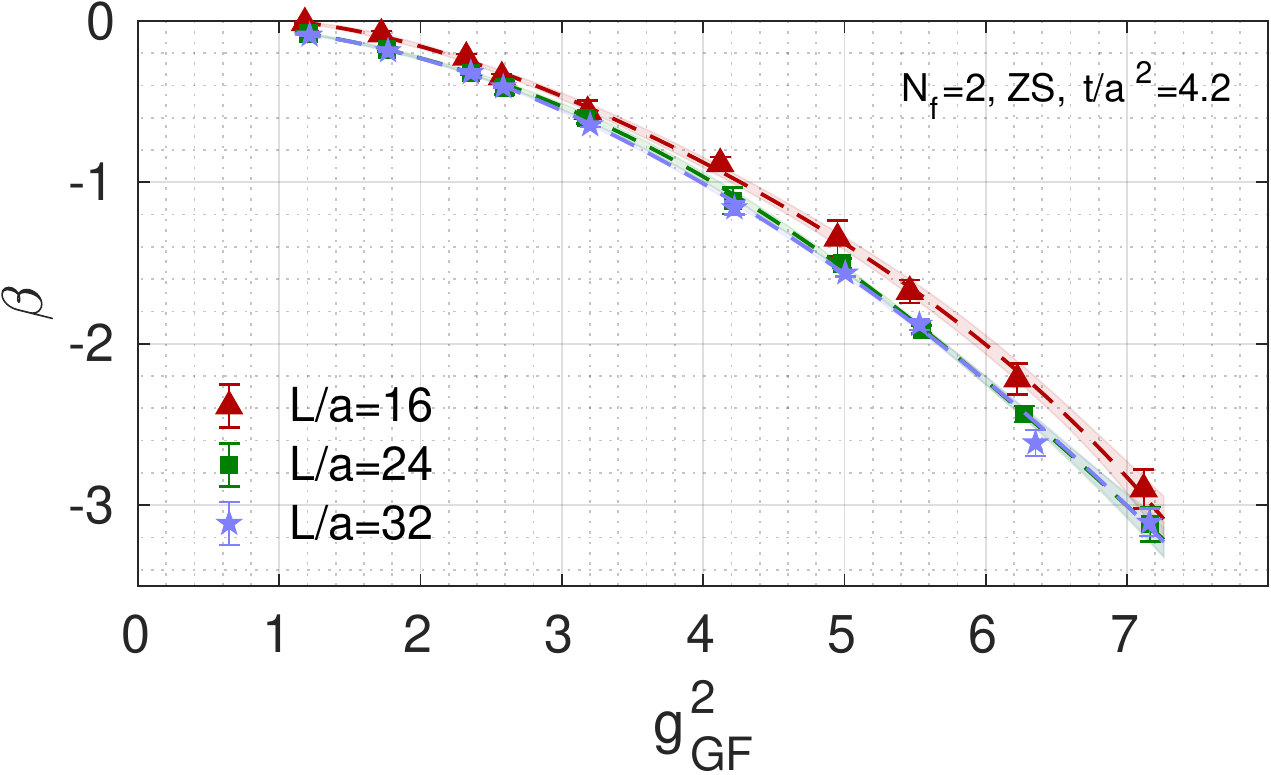}\\
\includegraphics[width=0.494\columnwidth]{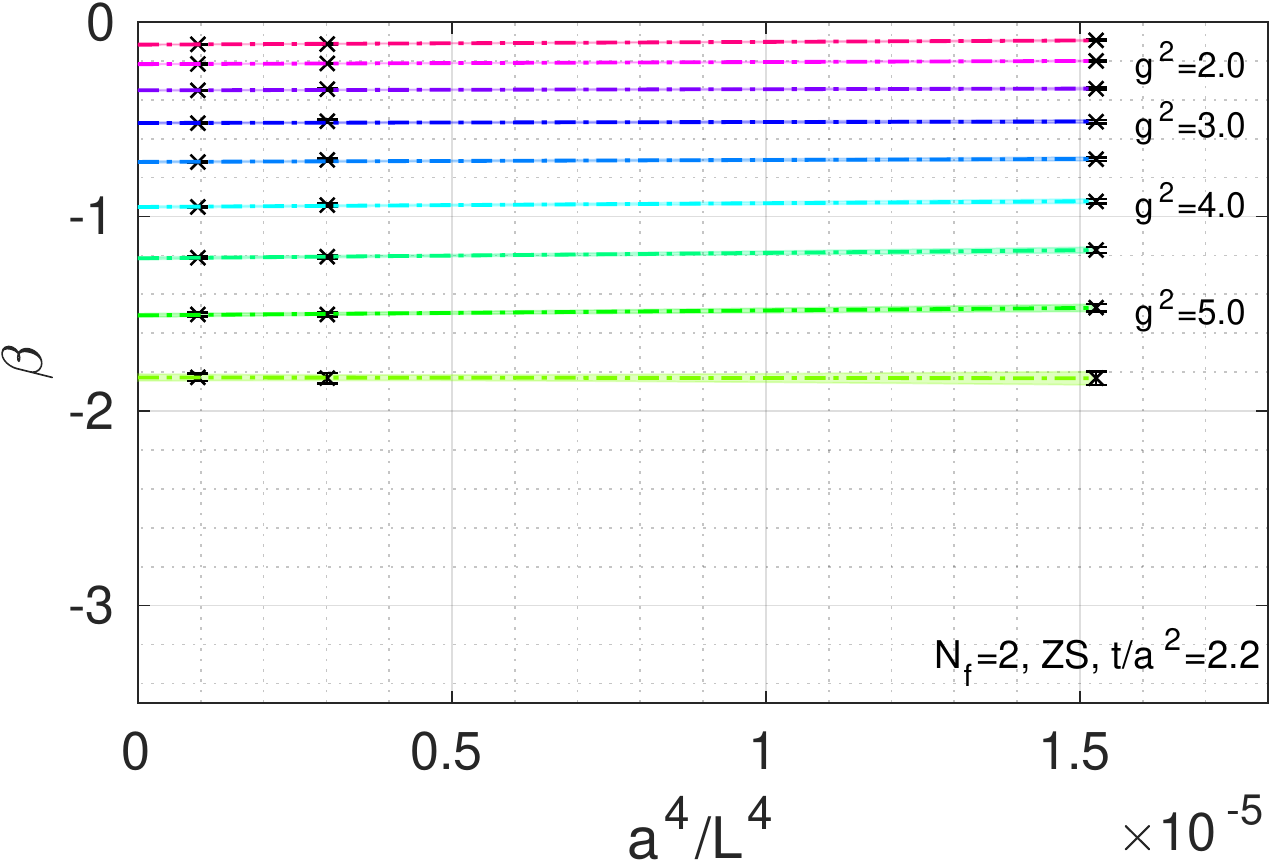} 
\includegraphics[width=0.494\columnwidth]{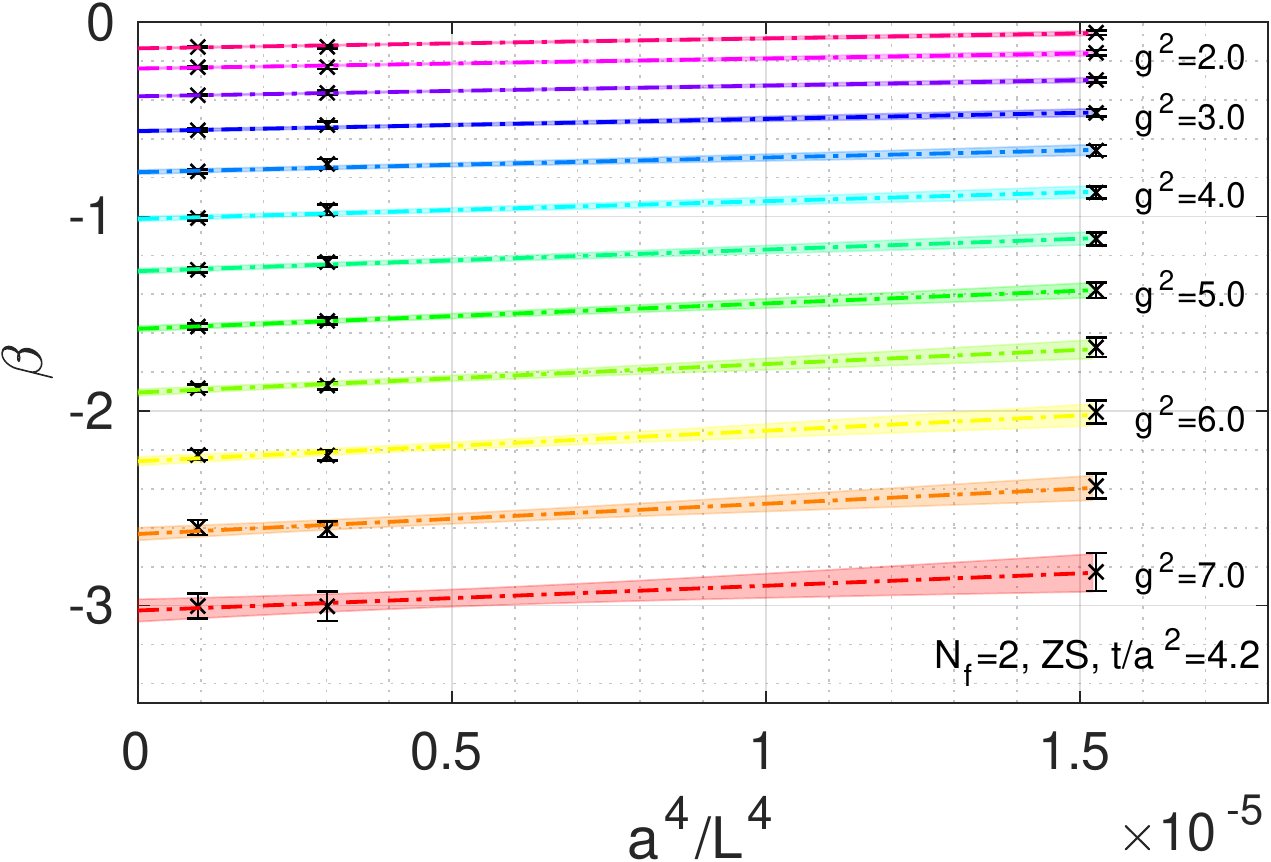}
\caption{ Top panels show the finite volume $\beta$ function at flow times $t/a^2=2.2$ and 4.2 ($a^2/t=0.455$ and $0.238$) for our three volumes for $N_f=2$. Dashed lines show a polynomial interpolation of the data points. Bottom panels present the infinite volume extrapolation at several $\gGF$ values for the same flow times.}
\label{fig:Linf_extrap_Nf2}
\end{figure}

\subsection{$N_f=2$}
We generate $16^3\times64$, $24^3\times64$ and $32^3\times64$ gauge field ensembles at 10 bare gauge coupling values ($\beta\equiv 6/g_0^2$ = 8.50, 7.00, 6.20, 6.00, 5.60, 5.20, 5.00, 4.90, 4.80, 4.70) using periodic boundary conditions in space, antiperiodic in time for the fermions. All ensembles are in the chirally symmetric regime, i.e. above the finite temperature phase transition. This choice allows us to run the simulations with $am=0$ and covers the coupling range $\gGF\lesssim 7.0$. 
\paragraph{A) Infinite volume extrapolation:}
The $L/a\to\infty$ limit has to be taken at fixed  $t/a^2$ and $g^2_{GF}$.  
The finite volume effects depend on $t/L^2$ and at leading order are proportional to $t^2/L^4$. We restrict the GF time in our analysis  such that the leading order contribution describes the data well. First we  determine  $\gGF(t)$ and its derivative on every ensemble, then interpolate $\beta(\gGF(t);L)$ for each lattice volume with a 4th order polynomial. This predicts the finite volume $\beta$ function as the function of the  renormalized coupling.  The top panels of Fig.~\ref{fig:Linf_extrap_Nf2} show both the lattice data and the interpolations at $t/a^2=2.2$ and $4.2$ ($a^2/t = 0.455$ and $0.238$) for the ZS combination. Using the predicted $\beta(\gGF)$ values we extrapolate in  $(a/L)^4$ to the infinite volume limit. The lower panels of Fig.~\ref{fig:Linf_extrap_Nf2} show this for several $\gGF$ at the same GF time as the top panels. We find that finite volume effects are negligible at small flow time and remain small even at $t/a^2=4.2$. As a consistency  check we compare extrapolations using all three volumes  to extrapolations using the two largest volumes only. While the errors of the infinite volume predictions change, the values are  consistent. Other flow and operator combinations show similar volume dependence.

\begin{figure}[tb]
\centering
\includegraphics[width=0.494\columnwidth]{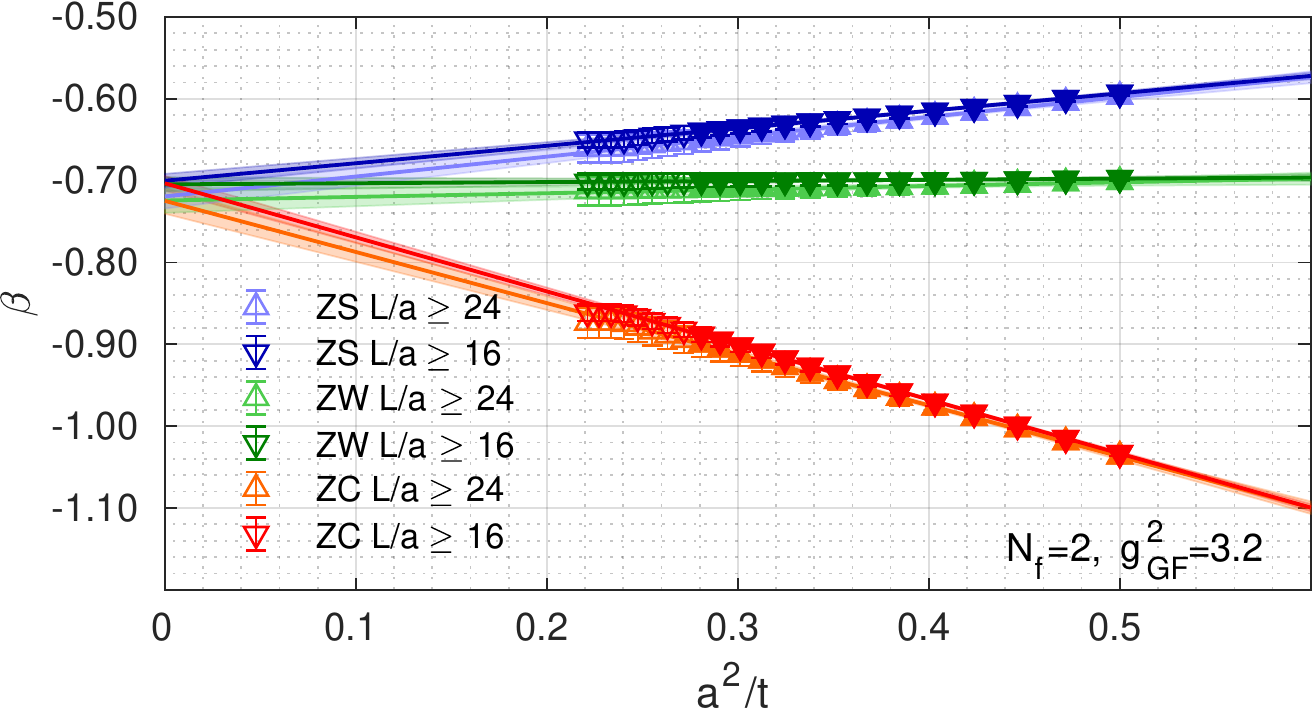}
\includegraphics[width=0.494\columnwidth]{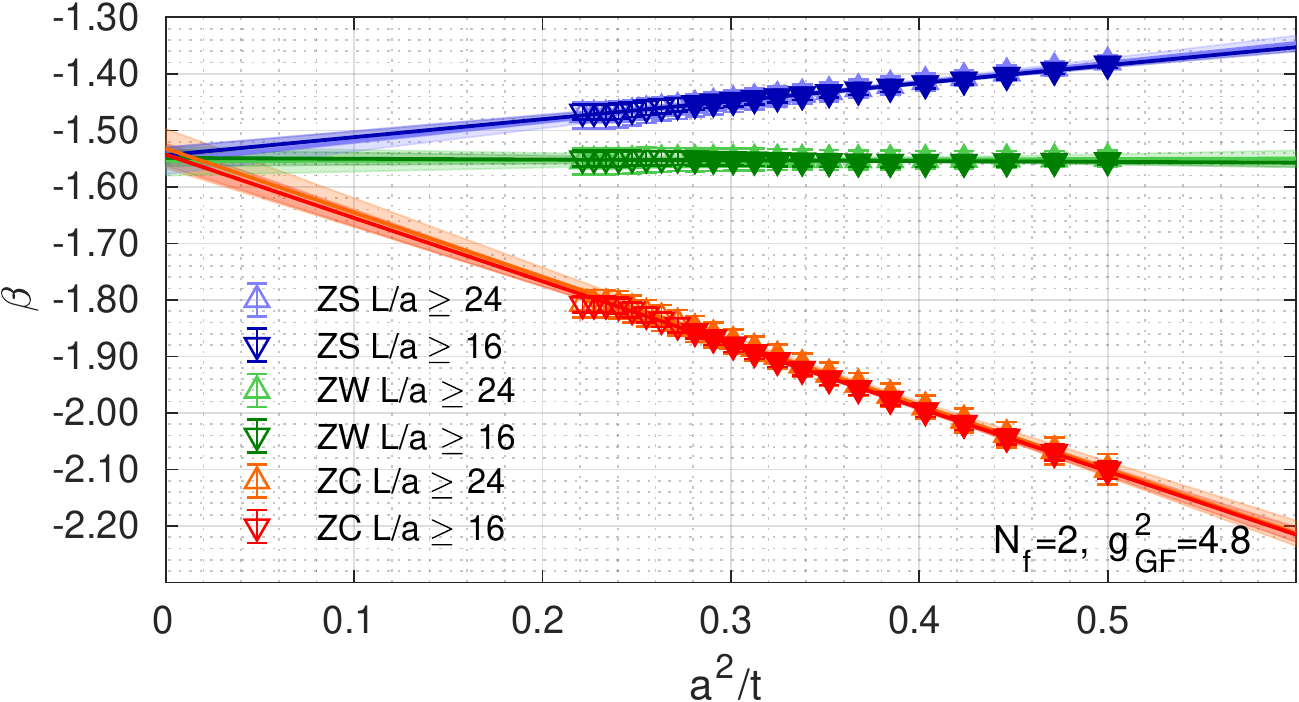}
\caption{Representative $a^2/t \to 0$ continuum limit extrapolation for $N_f=2$ at $\gGF=3.2$ (left panel) and  $\gGF=4.8$ (right panel). We show results for two infinite volume extrapolations and three operators, fitting filled symbols in the range $2.00 \le t/a^2 \le 3.64$ ($0.500\ge a^2/t\ge 0.274$). The (uncorrelated) fits are independent  but predict consistent $a^2/t=0$ continuum values.}
  \label{fig:cont_extrap_Nf2}
\end{figure}
\paragraph{B) Infinite flow time extrapolation: }
The continuum limit of the continuous $\beta$ function at fixed $\gGF$ is predicted in the $t/a^2 \to \infty$ limit. 
The range of $t/a^2$ values in the extrapolation has to be chosen with some care. The minimum flow time must  be large enough for the RG flow to reach the vicinity of the RT where all but at most one  irrelevant operators are suppressed.  The maximum flow  time is restricted by the requirement that the finite volume dependence follows the leading order $a^4/L^4$ dependence.  Any change of the continuum limit prediction  due to varying the minimal  or maximal flow time values can be incorporated as systematical  uncertainty. 

The functional form of the flow time dependence of $\beta(\gGF)$  is expected to be $(t/a^2)^{\alpha/2}$ where $\alpha < 0$ is the scaling dimension of the least irrelevant operator.   Around the GFP $\alpha=-2$  and we find that our data is well described by a linear $a^2/t$ dependence  for $t/a^2 \gtrsim 2.0$.    We show two examples of continuum extrapolation at $\gGF=3.2$ and 4.8 in Fig.~\ref{fig:cont_extrap_Nf2}. In both cases  we fit the data (filled symbols) in the range $2.00 \le t/a^2 \le 3.64$ ($0.500 \ge a^2/t \ge 0.274$). While the flow time $t$ is a continuous variable, in practice we evaluate $\gGF(t)$ with  finite step-size $\varepsilon=0.04$ and choose to dilute the data in $\Delta t=0.12$ intervals.We perform uncorrelated fits though correlations in $t$ could easily be accounted for in a bootstrap or jackknife analysis.  Once sufficiently large flow times are reached, the lower  flow times in the fit range impact only  the size of the uncertainties in the continuum limit  and the largest value of the coupling $g^2_{GF}$ which can be reached on a given data set. 
In Figure \ref{fig:cont_extrap_Nf2} we compare continuum limit extrapolations  obtained using Zeuthen flow  with Wilson  plaquette (ZW), Symanzik (ZS), and clover (ZC) operators.   We consider two different infinite volume extrapolations, using all three volumes or only the largest two. For illustration we show additional data at larger flow time using open symbols.  The  excellent agreement of the different extrapolations at the $a^2/t = 0$ limit is a  strong consistency check of the GF  time range and  the infinite volume extrapolation. 

It is worth  to point out that the Zeuthen flow, Wilson plaquette operator combination (green symbols in Fig.~\ref{fig:cont_extrap_Nf2}) show very little cut-off dependence and the data    are nearly constant in $a^2/t$. This is true  for both $\gGF$ shown in Fig.~\ref{fig:cont_extrap_Nf2} as well as other values we have investigated. The ZS combination also has relatively small cut-off  effects, but it is growing steadily as $\gGF$ increases.

\begin{figure}[tb]
\centering
\includegraphics[width=0.494\columnwidth]{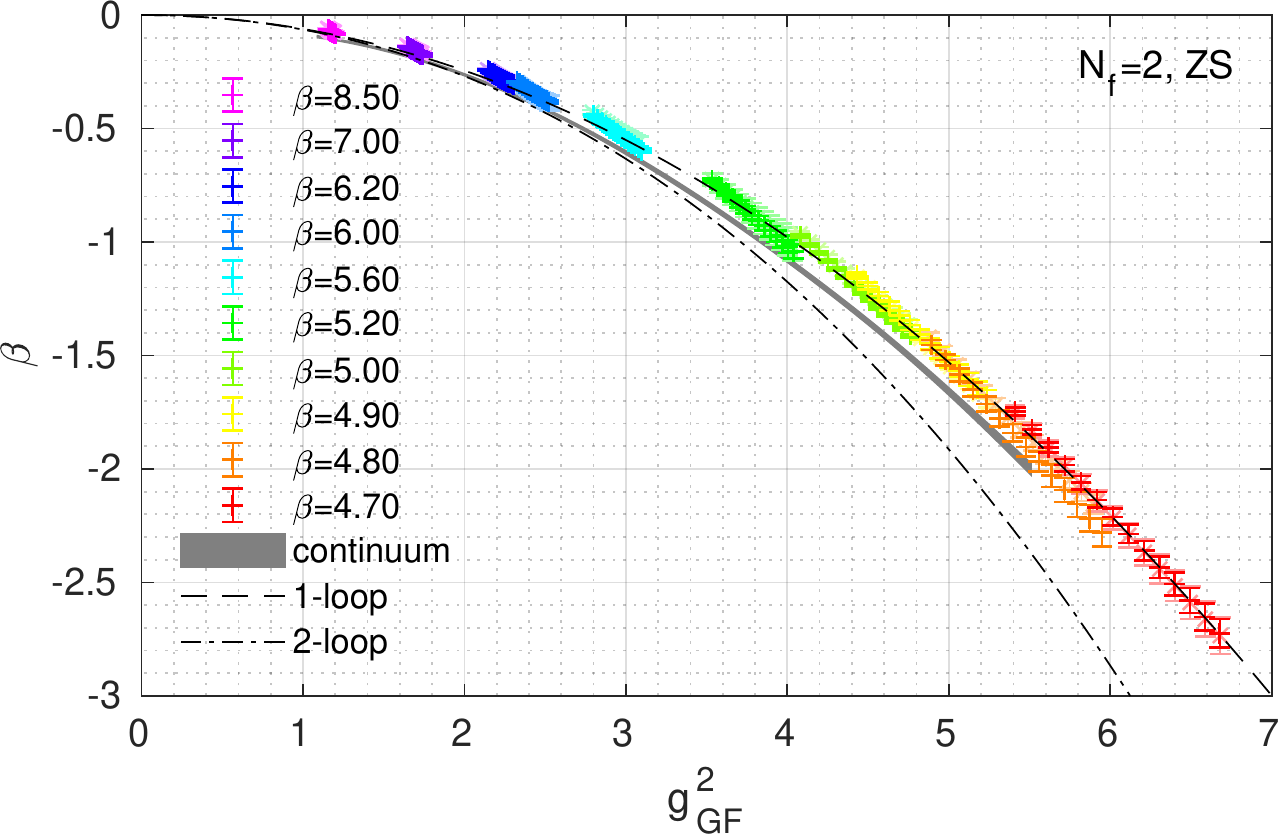}
\includegraphics[width=0.494\columnwidth]{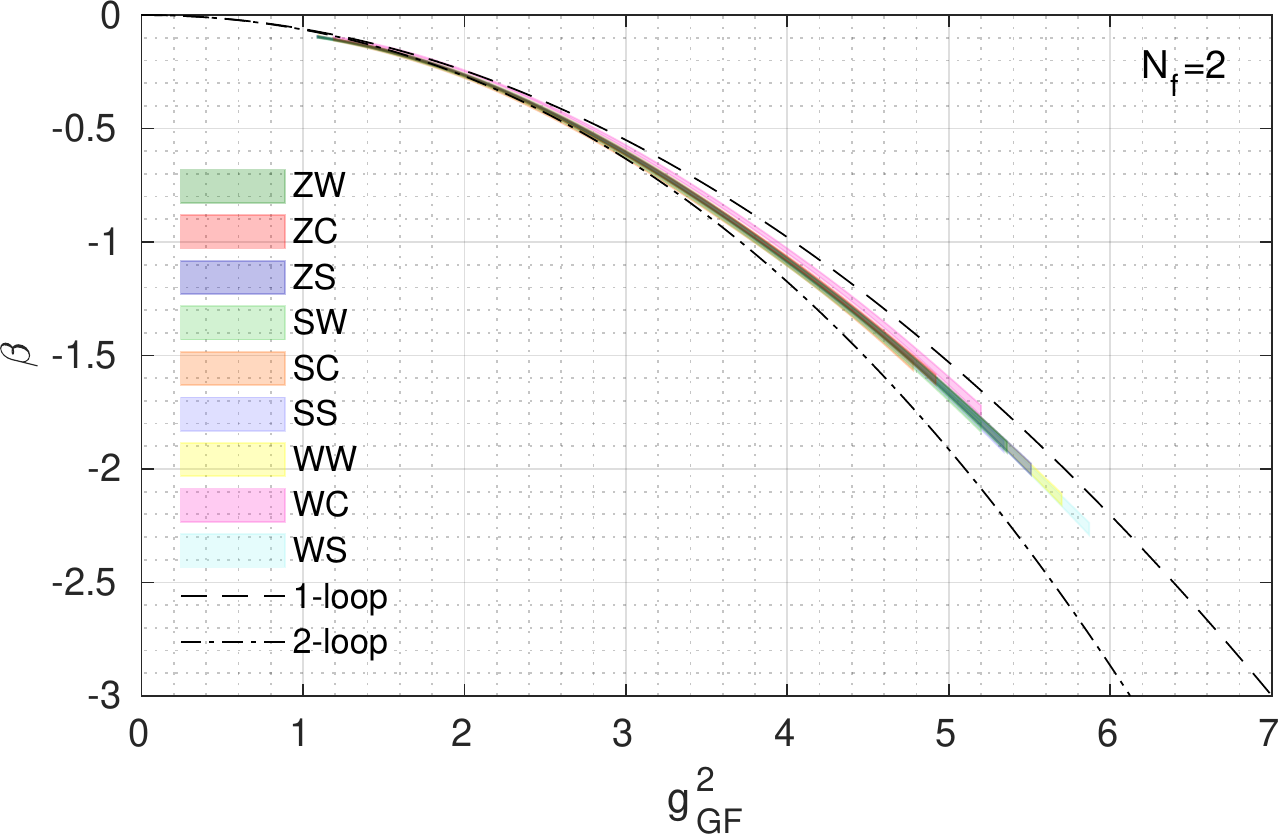}

\caption{ Left panel: Continuous RG $\beta$ function of 2-flavor QCD in the GF scheme. The grey band is the result of our full analysis with statistical uncertainties only. The colored data points show the lattice predictions using the ZS combination for  $32^3\times64$ ('$+$') and $24^3\times64$ ('$\times$') ensembles in a wide  range of bare couplings without any extrapolation or interpolation.  Only flow times $t/a^2 \in(2.0,3.64)$ are shown. The dashed and dash-dotted lines are  the perturbative 1- and 2-loop $\beta(g^2)$ functions. Right panel: Continuum limit of the continuous GF $\beta$ function predicted by nine different flow/operator combinations with statistical errors only. The different combinations are barely distinguishable and appear to be close to the 1-loop perturbative curve. }
\label{fig:beta-direct_Nf2}
\end{figure}

\paragraph{The continuous $\beta$ function:}
The Wilsonian RG description suggests that lattice simulations at a single bare coupling can predict, up to controllable cut-off corrections, a finite section of the RG $\beta$ function. In  practice, the finite lattice volume  limits the range where the infinite volume $\beta$ function is  well approximated. In the left panel of Figure \ref{fig:beta-direct_Nf2} the colored data points show the predictions for the RG $\beta$ function from   raw ZS  lattice data without infinite volume or continuum extrapolation. They trace out a single curve with overlapping predictions from different bare gauge couplings. The result of the full ZS analysis is shown by the gray band in Fig. \ref{fig:beta-direct_Nf2} which is in close agreement with the raw data.  The continuum limit predicted by different flow/operation combinations are consistent as shown in the right panel of Figure \ref{fig:beta-direct_Nf2}.
 In the $N_f=2$ system the coupling predicted by Zeuthen flow and Symanzik operator shows only small cut-off effects in the range of $\gGF\lesssim 6.0$ as is already evident from the continuum extrapolations shown in Fig.~\ref{fig:cont_extrap_Nf2}. The raw ZW lattice data show even smaller cut-off effects and completely overlap with the gray band of the full analysis. The continuous $\beta$ function approach predicts the running of the renormalized coupling in a transparent way where cut-off and finite volume effects are clearly  identifiable.

\subsection{$N_f=12$}
The analyses of the continuous $\beta$ function with 12 fundamental flavors closely follows the steps discussed above. The  12 flavor system has received a lot of attention but the predictions of its $\beta$ function are inconsistent\cite{Appelquist:2007hu,Hasenfratz:2011xn,Fodor:2011tu,Aoki:2012eq,Cheng:2013xha,Cheng:2014jba,daSilva:2015vna,Fodor:2016zil,Hasenfratz:2016dou,Fodor:2017gtj,Hasenfratz:2017qyr,Hasenfratz:2019dpr}.  The $\beta$ function is very small, so even small systematical errors can have a significant effect. At the same time, the RG structure can change suddenly when the conformal IRFP emerges as the sketch  in Fig.~\ref{fig:RG-flow} implies. The continuous $\beta$ function analysis is a new and partially independent approach that can shed light  on the origin of the controversies. 
 
 We use the ensembles  generated for the step scaling study published in Refs.~\cite{Hasenfratz:2017qyr,Hasenfratz:2019dpr}. These $(L/a)^4$ configurations  with $\beta\equiv 6/g_0^2$ =  4.15, 4.17, 4.20, 4.25, 4.30, 4.40, 4.50, 4.60, 4.70, 4.80, 5.00, 5.20, 5.50, 6.00, 6.50, and 7.00  have antiperiodic fermion boundary conditions in all four directions. In the analysis we only include the  $L/a=20$, 24, 28 and 32 sets, though in the infinite volume extrapolation we also show data from $L/a=16$ configurations.  From the preliminary analysis we show in the following only results using Zeuthen flow.

\paragraph{A) Infinite volume extrapolation:}
 \begin{figure}[tb]
\centering
\includegraphics[width=0.494\columnwidth]{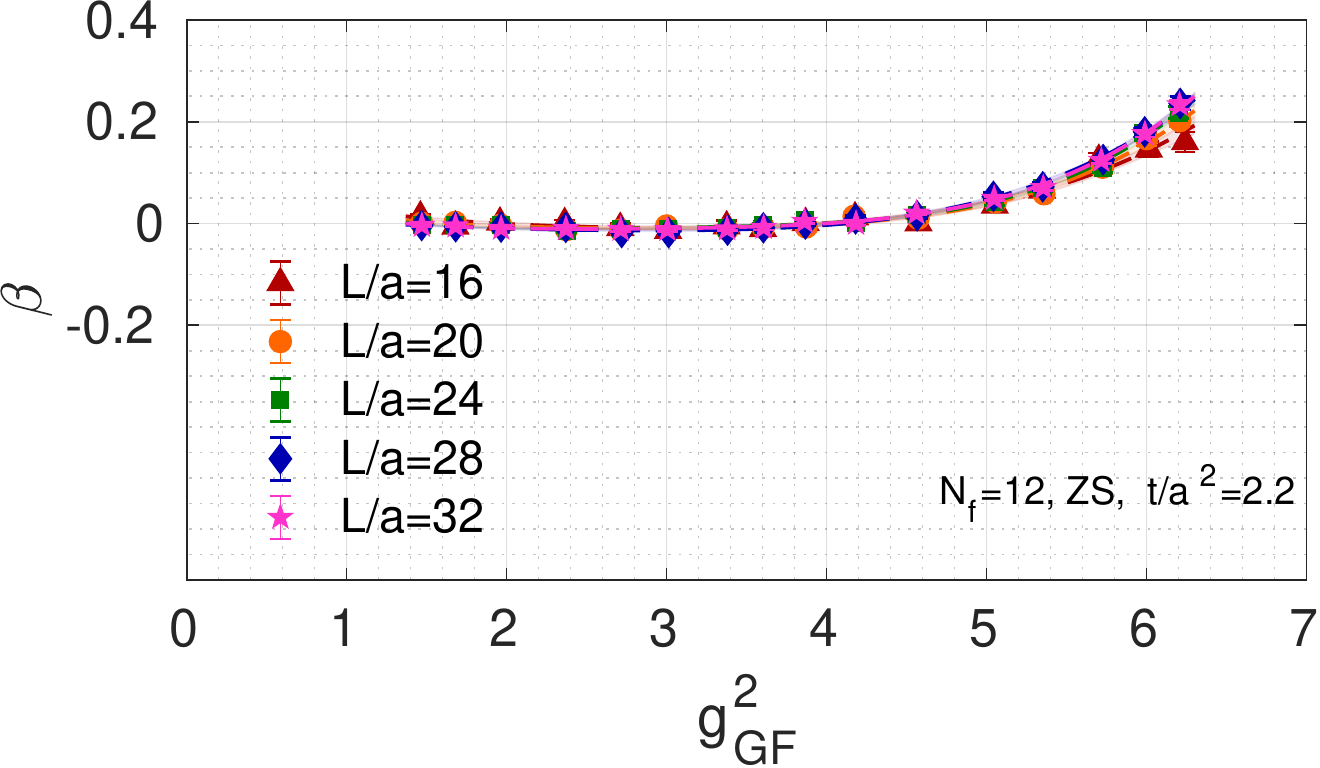}
\includegraphics[width=0.494\columnwidth]{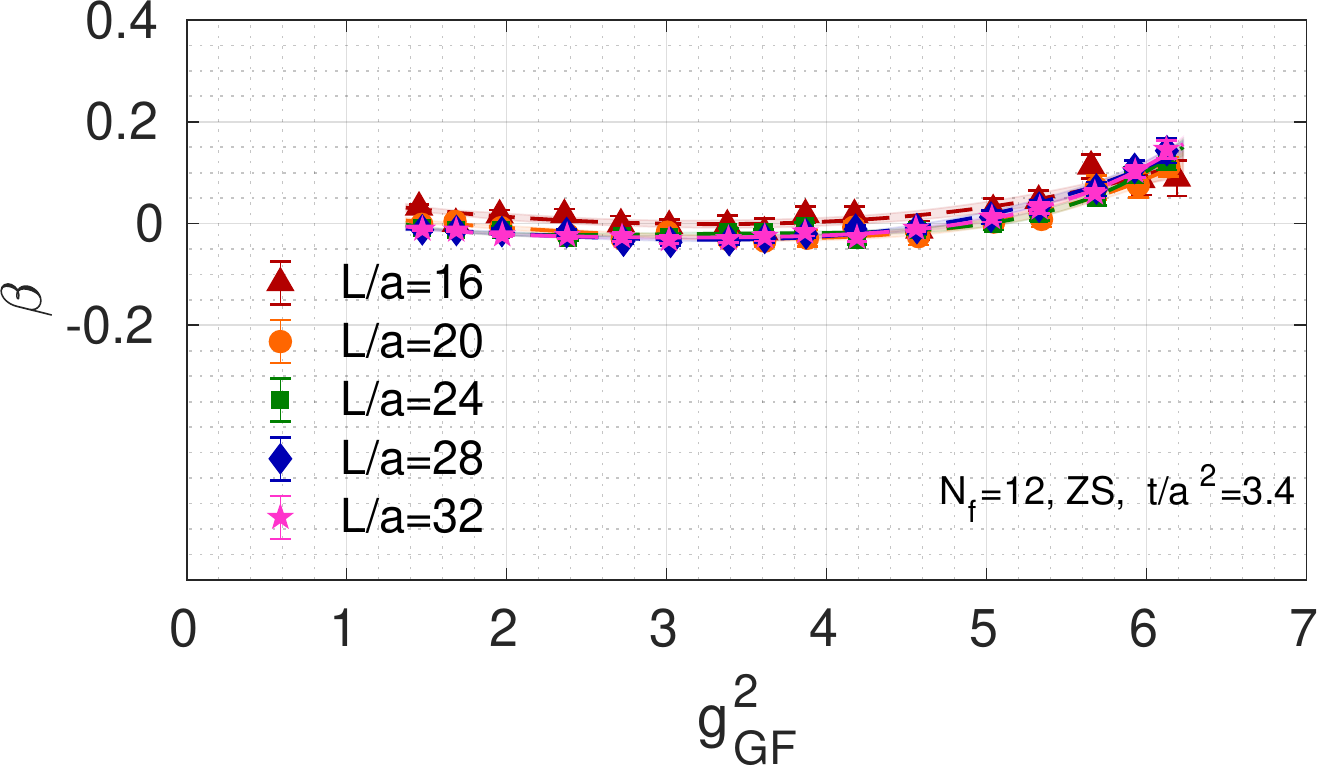}\\
\includegraphics[width=0.494\columnwidth]{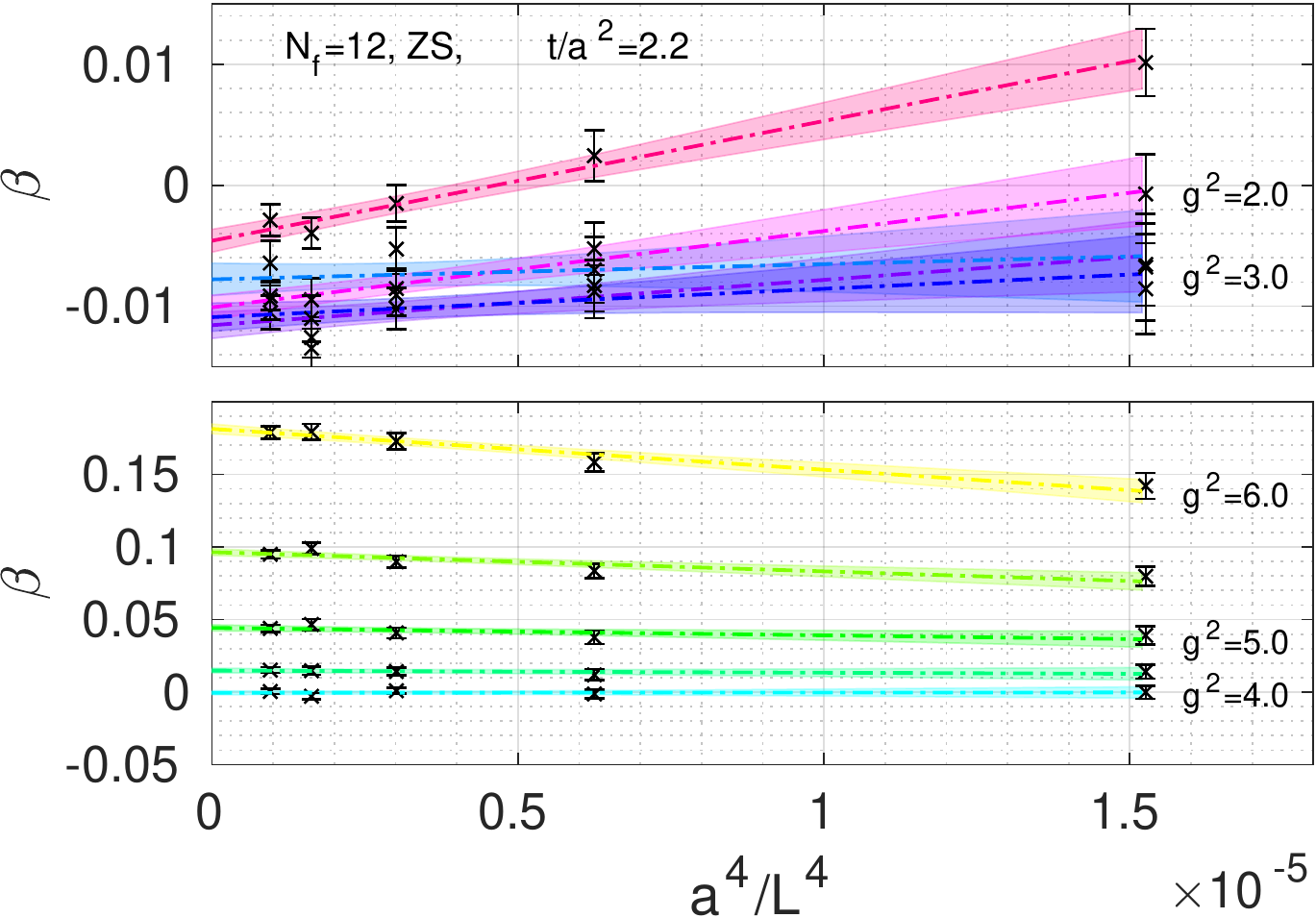} 
\includegraphics[width=0.494\columnwidth]{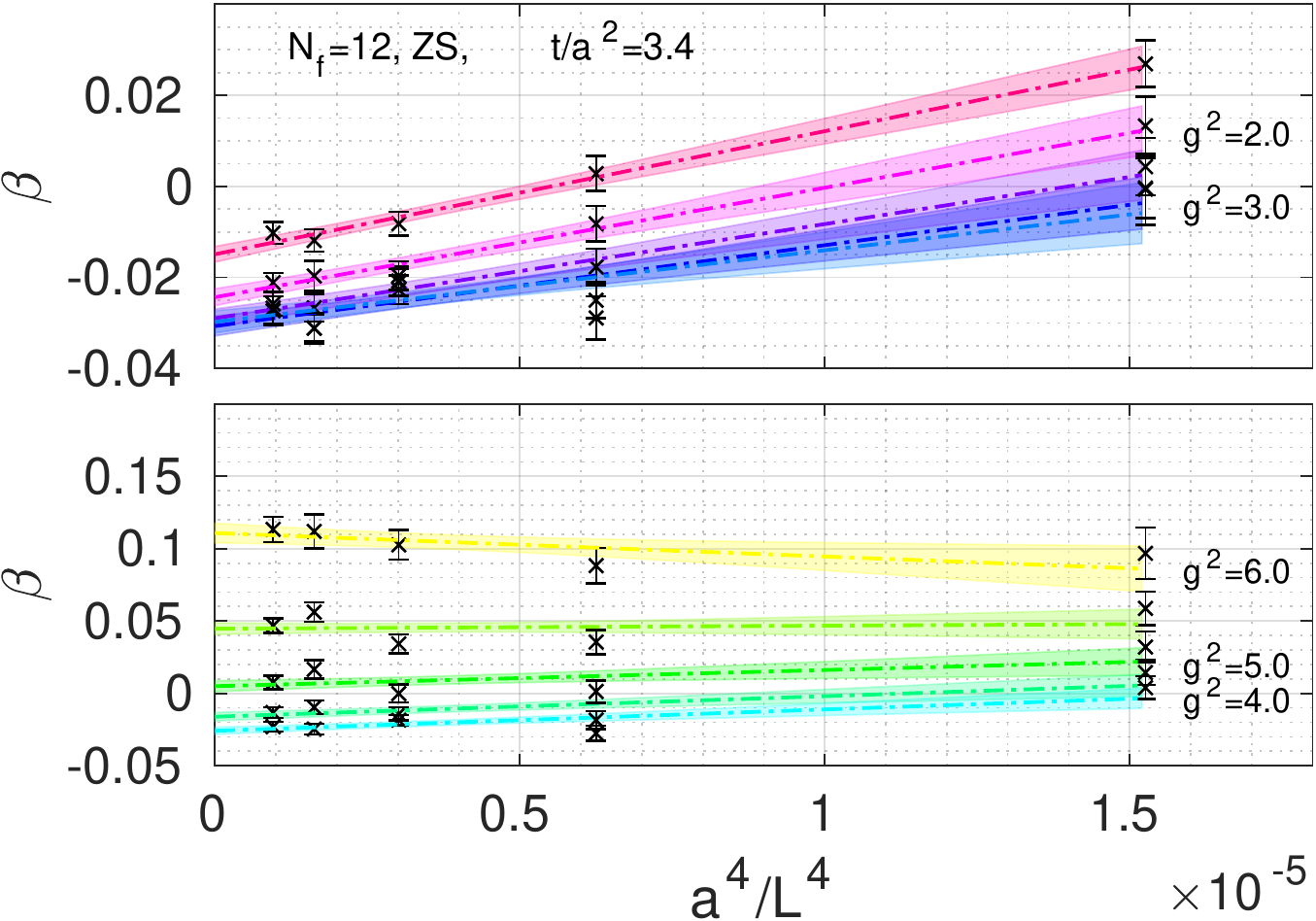}
\caption{The panels in the first row show the finite volume $\beta$ function at flow times $t/a^2=2.20$ and 3.40 ($a^2/t=0.455$ and $0.294$) for our five $N_f=12$ volumes. Dashed lines show a polynomial interpolation of the data points. The panels at the bottom present the infinite volume extrapolation at several $\gGF$ values for the same flow times. To resolve the small variations at weak coupling, the infinite volume extrapolations are shown in two panels. The upper ones have very small range allowing us to resolve a downward slope which however has only a tiny effect on the absolute value.}
\label{fig:Linf_extrap_Nf12}
\end{figure}
Figure \ref{fig:Linf_extrap_Nf12} shows details of the infinite volume extrapolation. Since the $\beta$ function is small, finite volume effects are more noticeable than in the $N_f=2$ study. Taking advantage of additional volumes, we consider the infinite volume limit restricted to $L/a\ge 24$ or $L/a\ge 20$. $L/a=16$ data are shown in  Fig.~\ref{fig:Linf_extrap_Nf12} only for further illustration of finite volume effects. Since the slowly varying $\beta$ function is small, we show the infinite volume extrapolations using two panels in Fig.~\ref{fig:Linf_extrap_Nf12}. The upper panel has an extremely small range in $\beta$ to show different extrapolations at weak coupling ($g_{GF}^2<4.0$), while the lower panel presents the stronger coupling range with a larger scale. Although the downward slope in the upper panels is resolved, the absolute variation is tiny.

\paragraph{B) Infinite flow time extrapolation:}
\begin{figure}[tb]
\centering
\includegraphics[width=0.494\columnwidth]{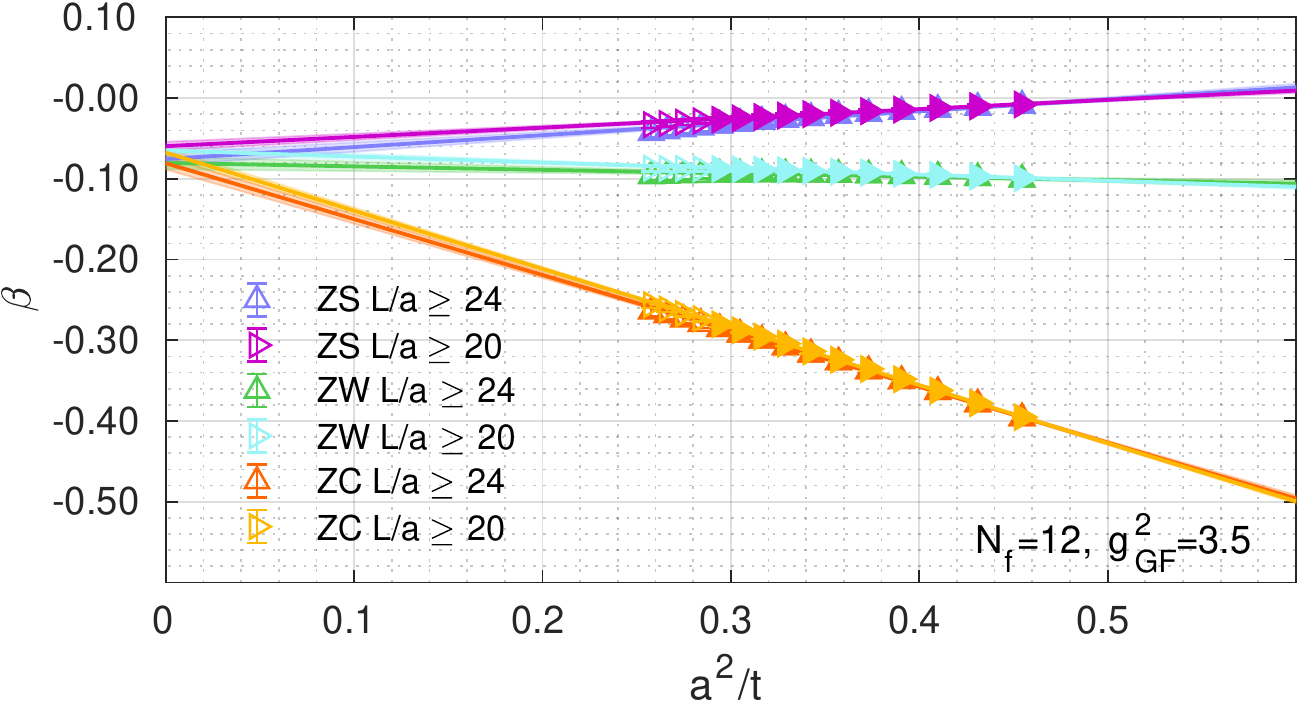}
\includegraphics[width=0.494\columnwidth]{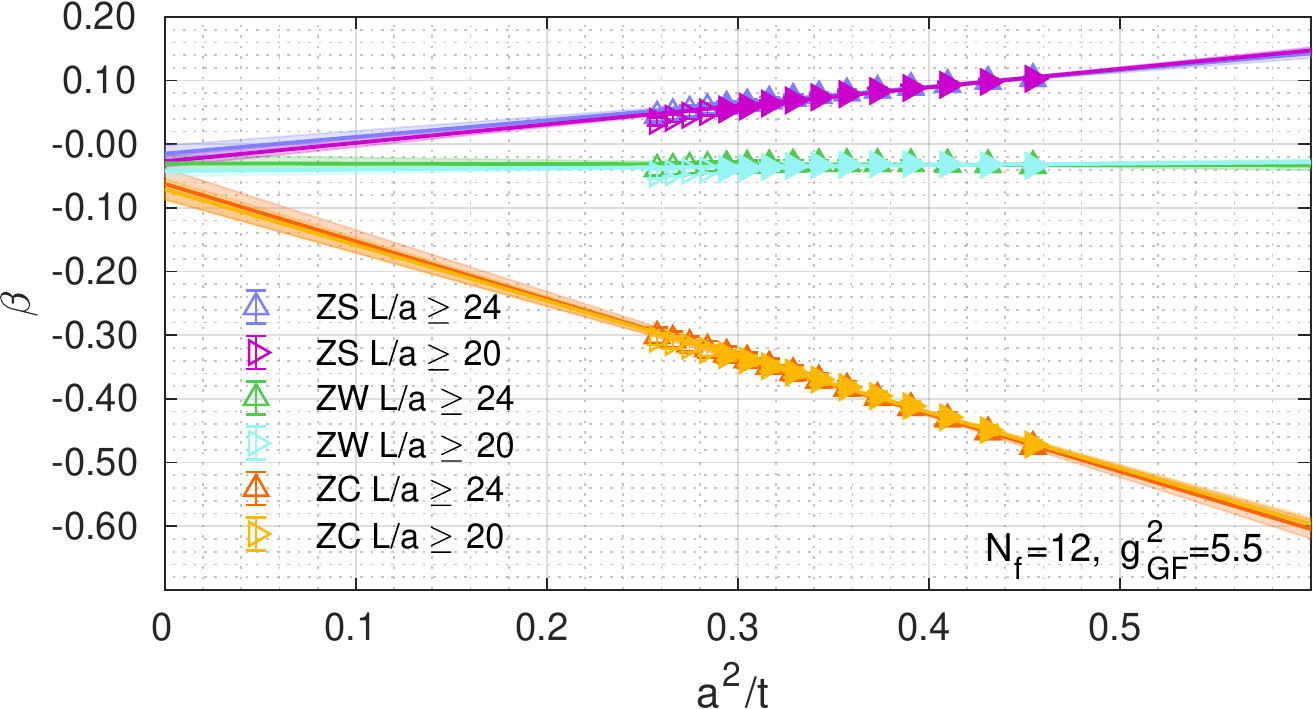}
\caption{Representative $a^2/t \to 0$ continuum limit extrapolation for $N_f=12$ at $\gGF=3.5$ (left panel) and $\gGF=5.5$ (right panel). We show results for two infinite volume extrapolations and three operators, fitting filled symbols in the range $2.20 \le t/a^2 \le 3.44$ ($0.455\ge a^2/t\ge 0.299$). The (uncorrelated) fits are independent  but predict consistent $a^2/t=0$ continuum values.}
  \label{fig:cont_extrap_Nf12}
\end{figure}
Figure \ref{fig:cont_extrap_Nf12} shows two examples of the continuum $a^2/t$ extrapolations. While the scaling exponent of the leading irrelevant operator is expected to change as we move away from the GFP, we cannot resolve such an effect. Fits  using the form $(t/a^2)^{\alpha/2}$ predict $\alpha \approx -2.0$ with  large uncertainties such that  the fit is consistent with $\alpha=-2.0$. Simultaneous  fits to two or three of the W, S, Z operators could provide sufficient  information to resolve  the scaling exponent. We will report on such an analysis in the future. Using a linear fit in the range $t \in (2.20,3.4)$, we predict consistent continuum limit values from W, S and C operators. 

Similar to the $N_f=2$ case, the Wilson operator shows the  smallest cutoff effects. The cyan/green data points are flat at both $\gGF$ values.  The clover operator has significantly larger cut-off corrections than the Wilson or Symanzik operators, rendering the continuum limit prediction from the ZC combination to be the least reliable of the three. 
\begin{figure}[tb]
\centering
\includegraphics[width=0.494\columnwidth]{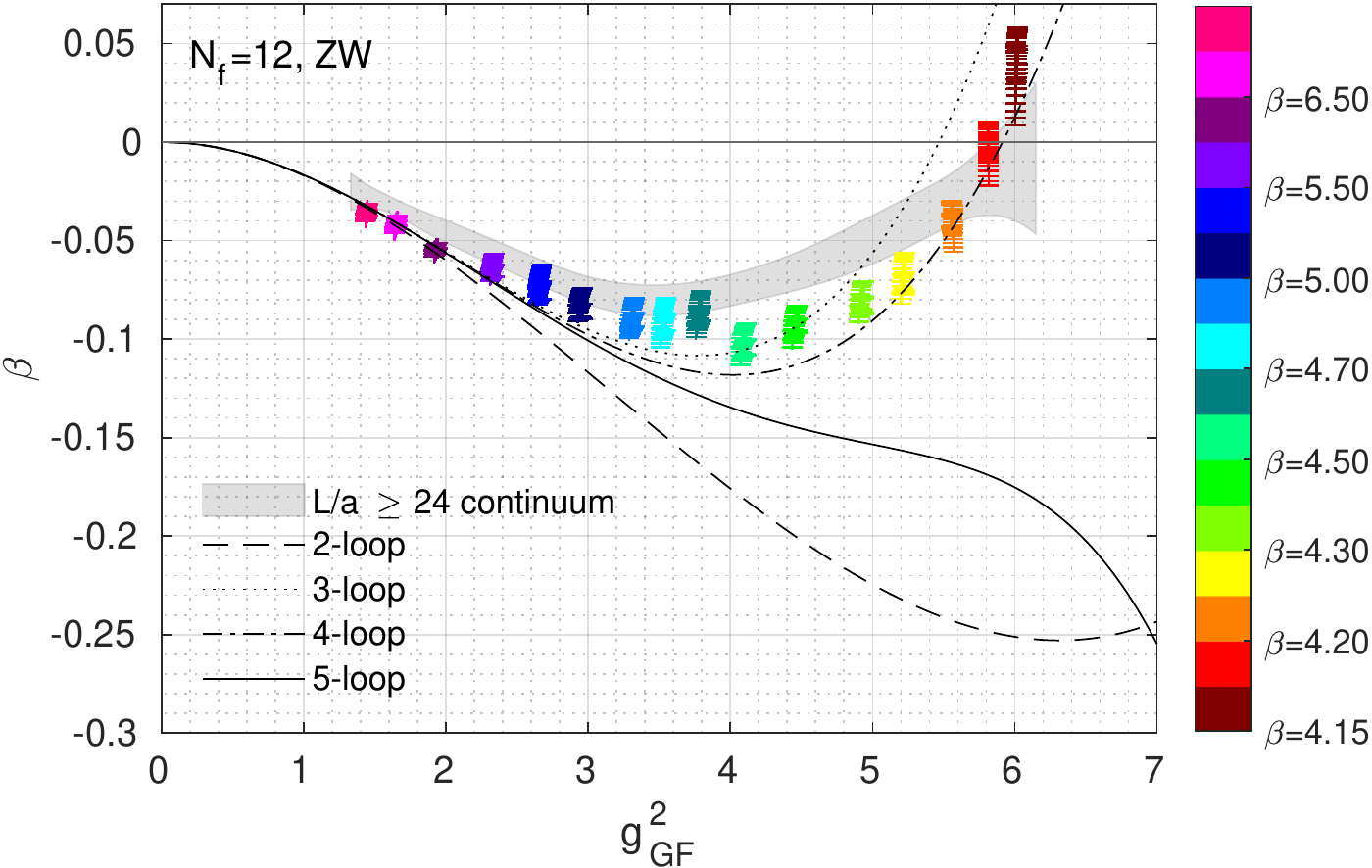}
\includegraphics[width=0.494\columnwidth]{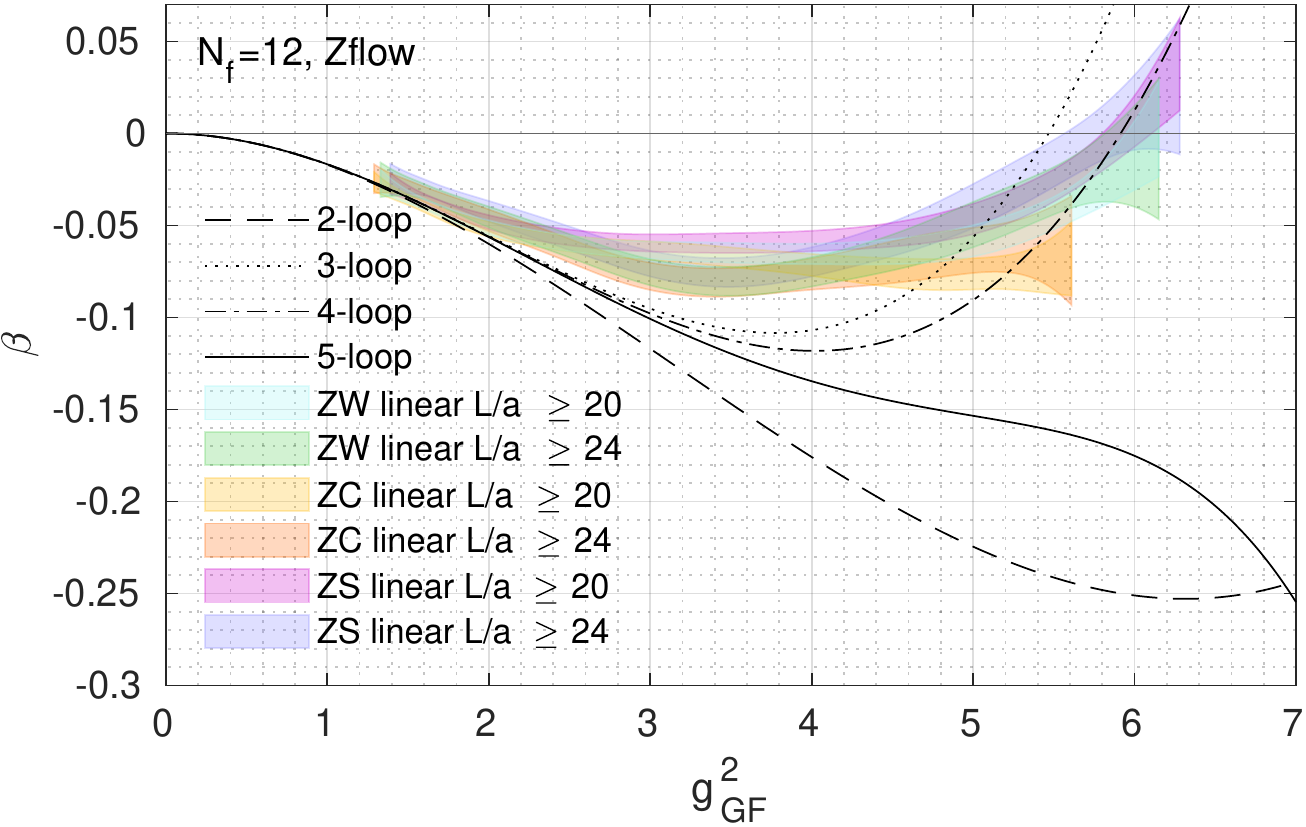}
\caption{ Continuous RG $\beta$ function of 12-flavor SU(3) model in the GF scheme using Zeuthen flow. The left panel shows the continuum limit obtained from the ZW combination using volumes with $L/a\ge 24$ overlayed with the raw ZW data (colored data points) from the $L/a=32$ ensembles. The colors indicate the 16 bare coupling values $\beta\equiv 6/g_0^2$ labeled by the colorbar on the right. In the panel on the right, we compare different continuum limit predictions using Zeuthen flow and the W, S, C operators as well as extrapolations based on $L/a\ge 20$ or 24. In addition to our nonperturbative results, we show predictions based on perturbation theory at 2-, 3-, 4-, and 5-loop order.} 
\label{fig:beta-direct_Nf12}
\end{figure}

\paragraph{The continuous $\beta$ function:}
Figure \ref{fig:beta-direct_Nf12} shows our predicted RG $\beta$ function based on Zeuthen flow. In the left panel we include the raw data from the ZW  flow/operator combination at GF time $t^2/a\in(2.20,3.4)$. The $\beta$ function of the $N_f=12$ system is  small, thus the renormalized $\gGF$ gauge coupling changes very slowly with the flow time. As a result the raw data in Fig.~\ref{fig:beta-direct_Nf12} explore a very small range in $g_{GF}^2$  at any given value of the bare coupling $\beta\equiv 6/g^2_0$. There is a more pronounced  fluctuation in $\beta(\gGF)$ visible mostly  due to the very small  scale of the plots.  As is predicted  in Fig.~\ref{fig:cont_extrap_Nf12}, the ZW data show only small cut-off effects resulting in  raw data which are very close to the continuum limit predictions. The ZS combination has larger cutoff corrections,  approaches the continuum predictions from above but exhibits also a somewhat larger reach in $g_{GF}^2$.  While the raw data of the 12-flavor system  does not offer the same description of the continuous $\beta$ function as we have seen in the 2-flavor case, it still offers intuition on how the continuum limit is approached. 

It is worth mentioning that the $\beta$ function predicted by the raw ZW flow increases with flow time up to $6/g^2_0\le4.25$ but changes direction, i.e. decreases towards more negative values for $6/g^2<4.25$. This qualitative change could indicate that  the influence of a possible nearby UVFP is getting strong on the RG flows.

In the right panel of Fig.~\ref{fig:beta-direct_Nf12}, we compare continuum predictions obtained using Zeuthen flow and the three different operators S, W, C. In general we observe very good agreement between the three flow/operator combinations shown, although ZC exhibits larger discretization errors as well as a shorter reach in $g_{GF}^2$ than ZS or ZW. In addition we check for finite volume effects by restricting the infinite volume extrapolations to volumes with $L/a \ge 20$ or $L/a\ge 24$. Again we observe that our results are consistent and we cannot resolve finite volume effects.

\section{Discussion}

We presented a method  based on a real-space RG  transformation with continuous scale change to determine the  continuous RG $\beta$  function. The validity of the approach relies on the nonperturbative Wilsonian RG transformations and is equally valid in the vicinity of the perturbative Gaussian FP, strongly coupled conformal IRFP or possible emerging UVFP both in conformal or infrared free systems.

 First we outlined the steps of determining the continuous $\beta$ function and validated the method  in 2-flavor QCD. Subsequently we followed the same steps to analyze existing GF data for the 12 flavor system. Results based on Zeuthen flow and Wilson plaquette, Symanzik and clover operators are consistent and predict an IRFP around $\gGF=6$. The value of the FP is scheme dependent. The continuous $\beta$ function corresponds to $c=0$ renormalization scheme, the value of the FP  is however similar to our determination in the $c=0.250$ gradient flow step-scaling scheme \cite{Hasenfratz:2019dpr}. 
 An advantage of the continuous RG transfromation is the possibility to resolve the scaling dimension of the irrelevant operators around a non-perturbative fixed point. In the $N_f=2$ system we found the scaling dimension $\alpha=-2.0$, consistent with the expectations around the GFP. At present we are not  able to resolve any difference  in the strong coupling regime. In the $N_f=12$ flavor system our preliminary analysis predicts $\alpha=-2.0$ with large errors which we expect to reduce in the future with a more sophisticated analysis. A similar method to determine the continuous $\beta$ function from lattice data generated at finite mass in chirally broken systems is discussed in Ref.~\cite{Fodor:2017die}. 

\section*{Acknowledgments}\vspace{-2mm}
We are very grateful to Peter Boyle, Guido Cossu, Anontin Portelli, and Azusa Yamaguchi who develop the \texttt{Grid} software library providing the basis of this work and who assisted us in installing and running \texttt{Grid} on different architectures and computing centers. A.H.~and O.W.~acknowledge support by DOE grant DE-SC0010005. We thank Slava Rychkov for correspondence and fruitful discussion that led to the phase diagram in the right panel of Fig.~\ref{fig:RG-flow}. We also thank Francesco Sannino for discussion on the 4-fermion system presented in Ref.~\cite{Rantaharju:2016jxy,Rantaharju:2017eej}.  
We are indebted to Daniel Nogradi for extending his original calculation of the finite volume correction factors on symmetric volumes to asymmetric lattices and sharing the result prior to publication. We thank Alberto Ramos for many enlightening discussions during the ``37th International Symposium on Lattice Field Theory'', Wuhan, China, and Rainer Sommer and Stefan Sint for helpful  comments.  We benefited from many  discussions with Thomas DeGrand, Ethan Neil, David Schaich, and Benjamin Svetitsky.
A.H.~would like to acknowledge the Mainz Institute for Theoretical Physics (MITP) of the Cluster of Excellence PRISMA+ (Project ID 39083149) for enabling us to complete a portion of this work.
O.W.~partial support by the Munich Institute for Astro- and Particle Physics (MIAPP) of the DFG cluster of excellence ``Origin and Structure of the Universe''.
Computations for this work were carried out in part on facilities of the USQCD Collaboration, which are funded by the Office of Science of the U.S.~Department of Energy and the RMACC Summit supercomputer \cite{UCsummit}, which is supported by the National Science Foundation (awards ACI-1532235 and ACI-1532236), the University of Colorado Boulder, and Colorado State University. This work used the Extreme Science and Engineering Discovery Environment (XSEDE), which is supported by National Science Foundation grant number ACI-1548562 \cite{xsede} through allocation TG-PHY180005 on the XSEDE resource \texttt{stampede2}.  This research also used resources of the National Energy Research Scientific Computing Center (NERSC), a U.S. Department of Energy Office of Science User Facility operated under Contract No. DE-AC02-05CH11231.  We thank  Fermilab,  Jefferson Lab, NERSC, the University of Colorado Boulder, TACC, the NSF, and the U.S.~DOE for providing the facilities essential for the completion of this work.

{\small
  \bibliography{../General/BSM}
  \bibliographystyle{JHEP-notitle}
}

%\begin{thebibliography}{99}
%\bibitem{...}
%....
%\end{thebibliography}

\end{document}